\documentclass{aa}
\usepackage{psfig}
\def\la{\;
\raise0.3ex\hbox{$<$\kern-0.75em\raise-1.1ex\hbox{$\sim$}}\; }
\def\ga{\;
\raise0.3ex\hbox{$>$\kern-0.75em\raise-1.1ex\hbox{$\sim$}}\; }
\newcommand{\zabs}{$z_{\rm abs}\,$}

\newcommand{\kms}{km~s$^{-1}\,$}
\newcommand{\cm}{cm$^{-2}\,$}

\begin{document}
\title{Extremely metal-poor  Lyman limit system at \zabs = 2.917
toward the quasar HE 0940--1050\thanks{Based on observations obtained at
the VLT Kueyen telescope (ESO, Paranal,
Chile), the ESO programme 65.O-0474(A) 
}
}
\author{S. A. Levshakov\inst{1}
\and I. I. Agafonova\inst{1} 
\and M. Centuri\'on\inst{2}
\and P. Molaro\inst{2}
}
\offprints{S. A. Levshakov}
\institute{
Department of Theoretical Astrophysics,
Ioffe Physico-Technical Institute,
194021 St. Petersburg, Russia
\and
Osservatorio Astronomico di Trieste, Via G. B. Tiepolo 11,
34131 Trieste, Italy
}
\date{Received 00  / Accepted 00 }
\abstract{We report on detailed Monte Carlo inversion 
analysis of the Lyman limit system observed at \zabs = 2.917 
in the VLT/UVES spectrum of the quasar HE~0940--1050.
Metal absorption lines of carbon and silicon in three ionization
stages and numerous atomic hydrogen lines have been
analyzed simultaneously. It is found that in order to match
the observations, the shape of the ultraviolet background
ionizing spectrum of Haardt \& Madau (1996) should be modified:
a spectrum with a higher intensity of the emission feature at
3 Ryd is required. 
It is also found that synthetic galactic
spectra (or different mixtures of them with power law spectra) 
cannot reproduce the
observations, indicating that the stellar contribution to the
ionizing background is negligible at $z \sim 3$.
For the first time a very low carbon abundance of
[C/H] = $-2.93\pm0.13$ and the abundance ratio 
[Si/C] = $0.35\pm0.15$ are directly measured 
in the Lyman limit system
with $N$(\ion{H}{i}) = $3.2\times10^{17}$ \cm.
If the absorber at \zabs = 2.917
provides an example of a pristine
gas enriched by the nucleosynthetic products of early generations of stars,
then the measured value of [Si/C]
seems to indicate that
the initial mass functions for these stellar populations
are constrained to intermediate masses, $M_{\rm up} \la 25M_\odot$. 
\keywords{Cosmology: observations --
Line: formation -- Line: profiles -- Galaxies:
abundances -- Quasars: absorption lines --
Quasars: individual: HE~0940--1050}
}
\authorrunning{Levshakov et al.}
\titlerunning{LLS at \zabs = 2.917 toward HE 0940--1050}
\maketitle

\section{Introduction}


Metal abundance measurements in quasar intervening
absorbers provide insight into
the chemical evolution of matter over cosmological time scales.
Since heavy elements were discovered in some Ly$\alpha$ 
clouds at  $z \sim 3$ by 
Cowie et al. (1995) and by Tytler et al. (1995), 
several mechanisms explaining
the origin of metals in the intergalactic medium (IGM) 
have been proposed. 
Nowadays it is supposed that at $z \ga  10$ the IGM
was pre-enriched by the first Population III (Pop III) 
stars formed from gas with zero 
metallicity (e.g. Bromm et al. 2001; Nakamura \& Umemura 2001) 
and enriched later due to 
disruption of low-mass protogalaxies (e.g. Madau et al. 2001)  
at $6 \la z \la 10$, and due to ejection of metals
from massive galaxies at $z \la 6$ (e.g. Aguirre et al. 2001; 
Scannapieco et al. 2002). 
It is clear that to see the signature of Pop~III stars one must go to
very low metallicities. The simulations of collapse and fragmentation of
primordial gas clouds suggest that in the first generation 
the stars may have been
very massive, $M_\ast \ga 100M_\odot$ (Abel et al. 2000; 
Bromm et al. 2001). However, the calculations by Nakamura \& Umemura (2001)
produced a bimodal initial mass function (IMF) for 
Pop~III stars with a second peak  
at $1-2M_\odot$. According to these calculations the first generation 
supernova ($10M_\odot < M_\ast < 35M_\odot$) and pair-instability supernova
($140M_\odot < M_\ast < 260M_\odot$) can produce metallicity in the range
$10^{-4}-10^{-3}Z_\odot$. 
Since the element yields and production
rate differ significantly for the massive ($10-35M_\odot$) 
and very massive
($140-260M_\odot$) stars (Woosley \& Weaver 1995; Umeda \& Nomoto 2001;
Heger \& Woosley 2002), the Pop~III IMF can be 
constrained by measuring the
abundance ratios in low metallicity cosmic objects.

In the present study we report on the discovery of extremely metal-poor
cloud at \zabs = 2.917 toward the quasar HE~0940--1050 which reveals
the lowest metallicity of [C/H]\footnote{Using the customary definition
[X/H] = log\,(X/H) -- log\,(X/H)$_\odot$. Photospheric solar abundances 
are taken from Holweger (2001).} 
$\simeq -3$ measured up to now at
high redshifts. All previously reported measurements of the metal
abundances in the Ly$\alpha$ clouds lie in the range [C/H] $> -2.4$
(Fan 1995; Songaila \& Cowie 1996; Levshakov et al. 2002a, hereafter LACM;
Levshakov et al. 2003, hereafter LADWD). 
We note that the lowest carbon abundance measured in Galactic
metal poor stars is [C/H] = $-3.6$ (Norris et al. 2001).

While dealing with metallicity measurements in high redshift absorbers,
one should take into account that the results are strongly dependent on
several assumptions on physical parameters poorly known such as 
geometry of the cloud, gas density distribution, shape and intensity of the
background ionizing radiation, ionizing mechanisms etc. Our results are
obtained with the Monte Carlo inversion (MCI) algorithms described in
detail in  Levshakov et al.(2000, hereafter LAK), LACM, LADWD.
Since this technique is relatively new, we briefly outline here its
basics. 

The main assumption is that all lines observed in an absorption system
arise in a {\it continuous} absorbing gas slab of a thickness $L$ with
a fluctuating gas density and a random velocity field. Numerous cosmological
hydrodynamical calculations performed in the previous decade have shown that 
the QSO absorption lines arise more likely in the smoothly fluctuating
intergalactic medium in a network of sheets, filaments, and halos
(e.g. Cen et al. 1994; Miralda-Escud\'e et al. 1996; Theuns et al. 1998).
This is also supported by modern high resolution spectroscopic
observations: the increasing spectral resolution reveals progressively
more and more complex profiles. Thus, such 
an approach seems to be more
physically justified as compared to the model of separate homogeneous clouds
which the commonly used procedure is based on.  
This procedure consists of the deconvolution of the complex
spectra into a set of separate Voigt profiles and the subsequent
estimation of ionization parameters (one for every subsystem)
from the obtained ionic column densities.
As shown in LAK and LACM, this simplified technique
may produce incorrect metallicities exceeding in some cases 1 dex.

Further we assume that within the absorber the metal abundances are
constant, the gas is in thermal and ionizing equilibrium and
it is optically thin for the ionizing UV radiation.
The intensity and the shape of the background ionizing radiation
is considered as an external parameter. 
Within the absorbing region the radial velocity $v(x)$ and the total
volumetric gas density $n_{\rm H}(x)$ along the 
line of sight are
considered as two continuous random functions which are represented
by their sampled values at equally spaced intervals $\Delta x$.
The computational procedure
is based on the adaptive simulated annealing (see LAK, LACM), the
fractional ionizations of different elements are computed 
at every space coordinate $x$
with the photoionization code CLOUDY (Ferland 1997). 

The MCI allows us
to recover self-consistently the physical parameters
of the intervening gas cloud; namely, the mean ionization parameter $U_0$,
the total hydrogen column density $N_{\rm H}$, 
the line-of-sight velocity and
density dispersions ($\sigma_{\rm v}$ and 
$\sigma_{\rm y}$, respectively) of the absorbing gas,
and the chemical abundances $Z_{\rm a}$ of all elements
involved in the analysis.
With these parameters we can further calculate 
the mean gas number density $n_0$,
the column densities for different species $N_{\rm a}$, 
the mean kinetic temperature
$T_{\rm kin}$, and the linear size $L$.
Having these comprehensive information we are able to classify the absorber
more reliably and hence to obtain important clues concerning the 
physical conditions in the intervening clouds.

The structure of the paper is as follows. Sect. 2 describes the data sets,
the estimated parameters for the \zabs = 2.917 Lyman limit system
are given in Sect. 3, the obtained results are discussed in Sect. 4, and
our conclusions are reported in Sect. 5.

\section{Observations and data reduction} 

Observations of the QSO HE~0940--1050 (B=16.6, z$_{\rm em}$ = 3.06) 
from the Hamburg/ESO survey
(Reimers et al. 1995; Reimers et al. 1996; 
Reimers \& Wisotzki 1997; Wisotzki et al. 2000)
have been obtained
with the Ultraviolet-Visual Echelle Spectrograph 
(UVES) on the  Nasmyth focus 
of the ESO 8.2m KUEYEN telescope, second unit of the VLT at Paranal, Chile.

The spectra were recorded with two different dichroic filters
which allow to use the UVES blue and red arms   
simultaneously as two independent spectrographs 
for both instrumental settings. 
One of the instrumental configurations covered the  wavelength ranges 
3300-3864 \AA\, and 4790-6816 \AA\,
with a wavelength gap between 5763 \AA\, and 5846 \AA, 
while the other covered the spectral regions
3740-4983 \AA\, and 6715-10400 \AA\, 
with a wavelength gap between 8530 \AA\, and 8677 \AA.

Three exposures of 3600 seconds each one of 
2700 seconds were obtained for each instrumental configuration
over the nights 26-29 March and 3 April 2000,  when the seeing, as given
by the telescope guide probe, was between 0.5  and 0.7 arcsec FWHM.

The slit widths were  
set at 1 arcsec and the CCDs were read-out in 2x2 pixel binned mode, 
resulting in 
a spectral resolution between 6.6 and 7.1 \kms. This  
full width at half maximum of 
the instrumental profile was measured from 
the widths of emission lines of the 
Thorium-Argon lamp used for the wavelength calibration of the spectra.    
     
The data reduction was performed using the ECHELLE context routines 
implemented in the ESO MIDAS package.     
Flat-fielding, cosmic ray removal, sky subtraction,  and
wavelength calibration 
were performed on each spectrum  separately.
Typical r.m.s. of the wavelength calibration is
$\leq$ 1 m\AA.    
  
The observed wavelength scale of each spectrum was then
transformed into vacuum, 
heliocentric wavelength scale. The single extracted spectra were then 
added together using weights proportional to their S/N.    
Finally, the local continuum was determined
in the average spectrum by using a spline to connect smoothly the  
regions free from absorption features. The continuum for the Ly$\alpha$ 
forest region
was fitted by using the small regions deemed to
be free of absorptions and by drawing 
an interpolating spline between them.
 
We confirm the identification of several metal absorption-line systems with
$z = 2.82$, 2.32, 1.918, and 1.06 performed from the low (4 \AA) resolution
spectra of HE~0940--1050 by Reimers et al. (1995). 
These systems will be analyzed in detail in a separate paper. 
A new Lyman limit system (LLS) with \zabs = 2.917, 
found in the high resolution ($\simeq 0.1$ \AA) VLT/UVES spectra,
is the subject of the present work.

\begin{figure}
\vspace{0.0cm}
\hspace{0.0cm}\psfig{figure=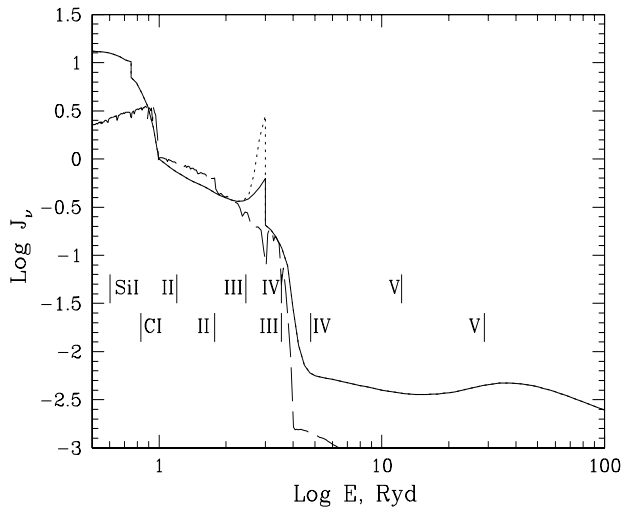,height=8.0cm,width=8.0cm}
\vspace{-0.8cm}
\caption[]{UV-background ionizing continua used in the present
calculations. The spectra have been normalized
so that $J_\nu$($h\nu=1$ Ryd) = 1.0. 
The solid curve shows the spectrum
computed by Haardt \& Madau (1996) at $z=3$. 
The dashed curve is a galactic spectrum from 
Leitherer et al. (1999) shown
in their Fig.~8(e) -- spectral energy distribution at 5 Myr with 
$Z/Z_\odot = 0.001$.
The dotted line at 3 Ryd shows the modification 
of the Haardt and Madau spectrum
required by the present observations. 
The positions of ionization thresholds
of different Si and C ions are indicated by tick marks 
}
\label{fig1}
\end{figure}

\section{Lyman limit system at \zabs = 2.917}

The Lyman limit system at \zabs = 2.9171 toward the quasar
HE~0940--1050
consists of  hydrogen Lyman series lines (from Ly$\alpha$ up to Ly-20) and
metal lines of
\ion{C}{ii} $\lambda1334$, 
\ion{Si}{ii} $\lambda1260$ and $\lambda1526$ (the \ion{Si}{ii} 
$\lambda\lambda1190, 1193$ lines are blended), 
\ion{Al}{ii} $\lambda1670$,
\ion{C}{iii} $\lambda977$, \ion{Si}{iii} $\lambda1206$,
\ion{N}{iii} $\lambda989$,
\ion{C}{iv} $\lambda1548$ and $\lambda1550$,  
\ion{Si}{iv} $\lambda1393$ and $\lambda1402$. 
The doublet \ion{O}{vi} $\lambda\lambda1031, 1037$ is
hopelessly blended and no conclusions about this important element can
be made.
Additionally the quasar spectrum beyond the
Lyman limit ($\lambda_{\rm obs} < 3572.4$ \AA) shows residual intensity
$I_\lambda \simeq 0.14$ giving us a model independent estimation of the
total neutral hydrogen column density of about $3\times10^{17}$ \cm.

\begin{table}
\centering
\caption{
Physical parameters of the \zabs = 2.917 Lyman limit system toward
HE~0940--1050 derived by the MCI procedure
}
\label{tbl-1}
\begin{tabular}{lc}
\hline
\noalign{\smallskip}
Parameter & \zabs = 2.9171 \\
\noalign{\smallskip}
\hline
\noalign{\smallskip}
Mean ionization parameter, $U_0$ & 7.6E-2$^c$ \\
Total H column density, $N_{\rm H}$, cm$^{-2}$ & 3.5E20$^c$ \\
Velocity dispersion, $\sigma_{\rm v}$, km~s$^{-1}$ & 50.0$^c$ \\
Density dispersion, $\sigma_{\rm y}$ & 1.5$^c$ \\
Chemical abundances$^a$: &  \\
$Z_{\rm C}$ & 4.6E-7$^c$ \\
$Z_{\rm N}$ & $<$ 5.3E-8 \\
$Z_{\rm Al}$ & 5.1E-9 \\
$Z_{\rm Si}$ & 9.0E-8$^c$ \\
$[Z_{\rm C}]$ & $-2.93\pm0.13$ \\
$[Z_{\rm N}]$ & $< -3.3$ \\
$[Z_{\rm Al}]^b$ & $-2.76\pm0.10$ \\
$[Z_{\rm Si}]$ & $-2.58\pm0.08$ \\
Column densities, cm$^{-2}$~: &  \\
$N$(H\,{\sc i}) & ($3.2\pm0.1)$E17 \\
$N$(C\,{\sc ii}) & ($4.8\pm0.3)$E12 \\
$N$(Al\,{\sc ii}) & 2.2E11 \\
$N$(Si\,{\sc ii}) & ($1.07\pm0.02)$E12 \\
$N$(C\,{\sc iii}) & ($1.20\pm0.06)$E14 \\
$N$(N\,{\sc iii}) & $<$ 1.4E13 \\
$N$(Si\,{\sc iii}) & ($1.35\pm0.02)$E13 \\
$N$(C\,{\sc iv}) & ($2.9\pm0.2)$E13 \\
$N$(Si\,{\sc iv}) & ($6.8\pm0.2)$E12 \\
Hydrogen number density, $n_0$, cm$^{-3}$ & 8.3E-4 \\
Mean kinetic temperature, K & 3.5E4 \\
Linear size, $L$, kpc & 140$^d$ \\
\noalign{\smallskip}
\hline
\noalign{\smallskip}
\multicolumn{2}{l}{
$^aZ_{\rm X}$ = X/H,\,
$[Z_{\rm X}] = \log (Z_{\rm X}) - \log (Z_{\rm X})_\odot$.}\\
\multicolumn{2}{l}{$^b$Al solar abundance from Grevesse \& Sauval (1998).}\\
\multicolumn{2}{l}{$^c$Internal error is 15\%.}\\ 
\multicolumn{2}{l}{$^d$Internal error is 20\%} 
\end{tabular}
\end{table}

\begin{figure*}
\vspace{0.0cm}
\hspace{0.0cm}\psfig{figure=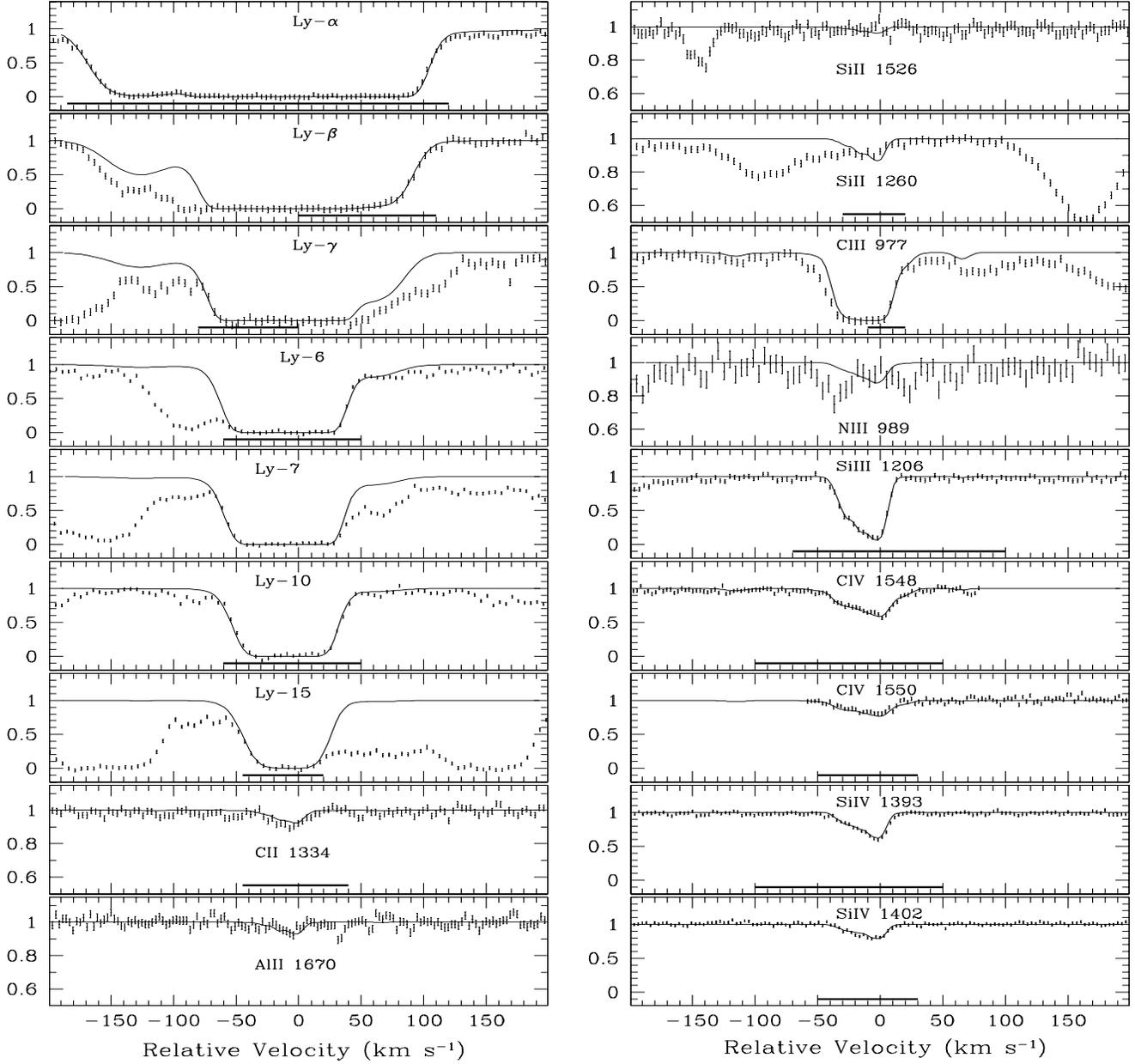,height=17.0cm,width=18.0cm}
\vspace{-0.5cm}
\caption[]{
Hydrogen and metal absorption lines associated with the
\zabs = 2.917 LLS toward HE~0940--1050
(normalized intensities are shown by dots with $1\sigma$ error bars).
The zero radial velocity is fixed at $z = 2.9171$. Smooth lines are
the synthetic spectra convolved with the corresponding point-spread
spectrograph function and computed with the physical
parameters listed in Table~1. Bold horizontal lines mark pixels included
in the optimization procedure. 
The normalized $\chi^2_{\rm min} = 0.97$ (the number of degrees of
freedom $\nu = 553$).
Profiles of Ly-$\beta$ ($\Delta v \la -70$ \kms),
Ly-$\gamma$ ($\Delta v \la -80$ \kms\, and $\Delta v \ga 40$ \kms),
Ly-7 ($\Delta v \ga 35$ \kms),
Ly-15 ($\Delta v \la -50$ \kms), and
\ion{C}{iii} $\lambda$977 ($\Delta v \la -35$ \kms)
are contaminated by other absorption lines (presumably from the
Ly-$\alpha$ forest).
The blue wing of Ly-6 ($\Delta v \la -60$ \kms) is blended by the
Ly-$\beta$ line from the \zabs = 2.5536 system. The red wing of Ly-15
($\Delta v \ga 20$ \kms) is blended by the Ly-$\beta$ and Ly-5 lines
from, respectively, the \zabs = 2.4962 and 2.8237 systems
(the systems with \zabs = 2.4962 and 2.5536 are identified in
I. I. Agafonova et al. 2003, in preparation) 
}
\label{fig2}
\end{figure*}

\begin{figure*}
\vspace{0.0cm}
\hspace{0.0cm}\psfig{figure=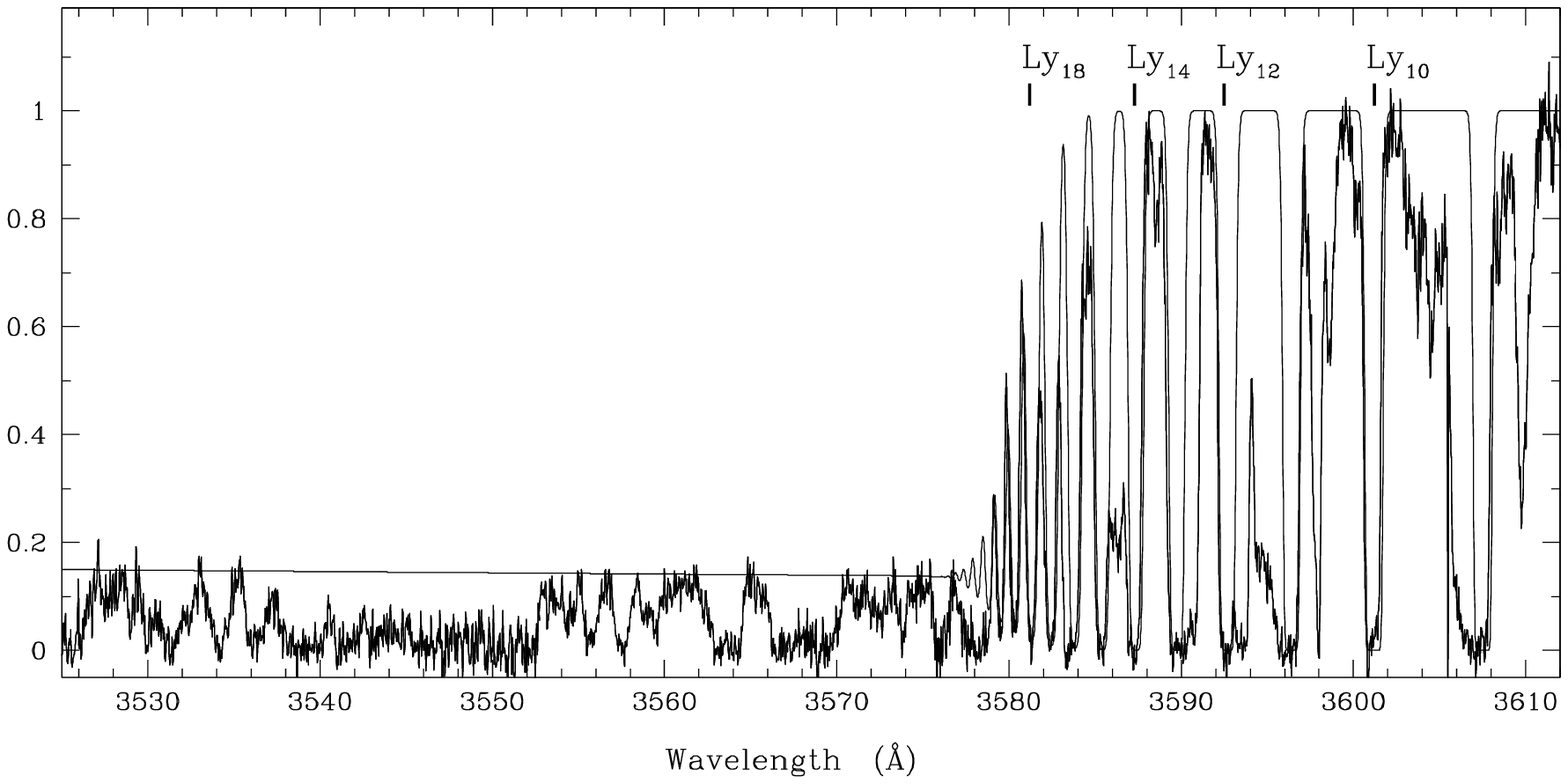,height=10.0cm,width=18.0cm}
\vspace{-3.8cm}
\caption[]{
Portion of the VLT/UVES spectrum of HE~0940--1050 in the region of the
Lyman limit absorption at \zabs = 2.917. The panel illustrates
the higher order hydrogen Lyman series lines from the
\zabs = 2.917 LLS. Smooth curve represents
a synthetic spectrum calculated with the 
physical parameters listed in Table~1.
Continuum depletion between 3538 \AA\, and
3555 \AA\, is due to the Ly$\alpha$ absorption from the
damped Ly$\alpha$ system at \zabs = 1.918
}
\label{fig3}
\end{figure*}

The first attempts to inverse the system under study were made with the
HM ionizing spectrum corresponding to $z = 3$. In spite of many trials
with different initial conditions it turned out to be impossible to gain
a good fitting for all available lines: the \ion{Si}{ii} $\lambda1260$
line was systematically overestimated, whereas the \ion{Si}{iv}
$\lambda1393, 1402$ doublet -- underestimated. This was the reason why
we started to modify the shape of the background ionizing spectrum.
After numerous experiments with different power law spectra, galactic
spectra from Leitherer et al. (1999) and combinations of them it was
found empirically that the self-consistent fitting of all observed
lines can be achieved if we adopt the HM spectrum with about 4 times
enhanced intensity between 2 and 3 Ryd (shown by the
dotted line in Fig.~1). Can this emission peak be present in the
real intergalactic UV background ? Yes, indeed.
The bump at 3 Ryd in the HM spectrum is due to \ion{He}{ii} Ly$\alpha$
recombination radiation produced by the photoionized Ly$\alpha$ forest
clouds. The primary quasar UV continuum spectra were adopted
by HM in the form of power law $\nu^{-1.5}$ (for $\lambda < 1216$
\AA). 
But it is known that the far UV region of quasar radiation
sometimes contains rather strong emission lines.
For instance, quasar 
\ion{He}{ii} Ly$\alpha$ emission at 3 Ryd was observed in
spectra of HS~1700+6416 (Davidsen et al. 1996; Reimers et al. 1998),
Q~0302--003 (Jakobsen et al. 1994; Heap et al. 2000),
and in the composite QSO spectrum (Telfer et al. 2002),
the last spectrum also shows
additional three wide emission features
between 304 \AA\, and 500 \AA.  
Thus the strong bump at 3 Ryd can be explained qualitatively. 
However, to estimate its exact shape and the
amplitude, calculations accounting for the presence of emission lines
in far UV QSO spectra and for radiative transfer effects in an
inhomogeneous universe are needed.
From the analysis of the \zabs = 2.917
system we only conclude that to match all Si and C 
lines, about 4 times more photons with energies between 
2 and 3 Ryd are required
(the dotted line shown in Fig.~1 is only one possible example, 
the emission bump at 3 Ryd may be broader and lower as well).

Due to the large number of ionic
transitions in the \zabs = 2.917 LLS and 
due to the very high S/N ratio this system
is a sensitive probe of the shape of the background photoionizing
radiation. For instance,
it was impossible to describe self-consistently all
observed profiles of hydrogen and metal absorption lines
using any of the composite UV spectra ($\lambda < 912$ \AA)
of young galaxies from Leitherer et al. (1999)
(a typical galactic spectrum is shown in Fig.~1 by the dashed line)
or any of the combined spectra calculated by mixing (with different
weights) a spectrum of the young galaxy with 
the HM UV background. 
From this, we do not 
confirm the result by Steidel et al. (2001) 
that the Lyman-break galaxies produce a dominated photoionizing
background at $z \sim 3$. 

The results obtained with the MCI and the modified HM spectrum (Fig.~1)
are presented in Table~1 and illustrated in
Figs.~2-4 (for details of calculations see LAK and LACM).
Parts of line profiles included in the least-squares minimization
are marked by horizontal lines in each panel in Fig.~2.
We also tried to probe the presence of deuterium absorption but found that
the wide Ly$\alpha$ profile is not sensitive to any additional D absorption.
The blue wing of the Ly$\beta$ line is strongly blended and this line 
cannot be used in the D/H estimations.

Fig.~2 shows that the observed profiles 
(portions free from blending)
are well represented.
A self-consistent fitting of all available lines has revealed that
the blue wing of the \ion{C}{iii}\,$\lambda 977$ line is partly
contaminated by some Ly$\alpha$ forest absorption. 

The profiles of the \ion{Al}{ii}\,$\lambda 1670$ and
\ion{N}{iii}\,$\lambda 989$ were
calculated 
in a second round using the velocity 
$v(x)$ and density $n_{\rm H}(x)$
distributions already obtained
and the metallicities chosen in such a way that the synthetic
spectra did not exceed 1 $\sigma$ deviations from the observed intensities. 

The Lyman series lines higher than Ly-12 are lying on the right wing of the
damped Ly$\alpha$ (DLA) identified at \zabs = 1.918  
(Fig.~3). Using the neutral hydrogen column density in the
\zabs = 2.917 LLS, we estimated the neutral hydrogen column density
for the DLA system of
$N$(\ion{H}{i}) $\simeq 1.0\times10^{20}$ \cm\,
(this DLA system is described in detail in 
M. Centuri\'on et al. 2003, in preparation).

Fig.~4 illustrates the distributions of the gas density and the radial
velocity inside the \zabs = 2.917 LLS. It is clearly seen that
the co-existence of lowly and highly 
ionized species in this system is due to density fluctuations.
A wide range between $x = 0.15$ and 0.37 with very low gas density is
also responsible for the blue-side asymmetry seen in the metal line
profiles in Fig.~2.

\section{Discussion}

According to the recovered parameters, 
the LLS at  \zabs = 2.917 is a very large could ($L > 100$ kpc) with very
low metal content.
What could it be related to ?

In our code  $L$ is calculated from the ratio $L = N_{\rm H}/n_0$, 
and $n_0 \propto (1+\sigma^2_{\rm y}) J_{912} / U_0$ 
(here $J_{912}$ is
the Lyman-limit specific flux in \, erg \cm\,s$^{-1}$ Hz$^{-1}$ sr$^{-1}$).
Thus the intensity of the background radiation 
can  influence the mean gas density and, hence, the absorber size. 
We used in our calculations $J_{912} = 0.4\times10^{-21}$\,
erg \cm\,s$^{-1}$ Hz$^{-1}$ sr$^{-1}$\, given by 
Haardt \& Madau (1996) for the
mean background flux at $z = 3.0$.
This value is poorly known and may 
be affected by local sources. 
But since we have no clues on how this value might be 
changed for this particular   
case, we will rely on the estimated linear size. 

For linear sizes as large as hundreds of kpc 
the velocity differences due to the
Hubble expansion, $\Delta v_{\rm H}$, become 
already noticeable. At \zabs = 2.917 
and $L = 140$ kpc
the velocity difference is $\Delta v_{\rm H} = 81$ \kms, 
assuming $\Omega = 1$ and the present day Hubble 
constant $H_0 = 75$ \kms\, Mpc$^{-1}$.
The corresponding
velocity dispersion caused by the Hubble flow is then
$\sigma_{\rm H} = \Delta v_{\rm H}/\sqrt{12} = 23$ \kms. 
The velocity dispersion 
of the bulk material deduced by
the MCI is $\sigma_{\rm v} = 50$ \kms, 
implying that the 
line-of-sight peculiar velocity dispersion is
about 44 \kms.

\begin{figure}
\vspace{0.0cm}
\hspace{0.0cm}\psfig{figure=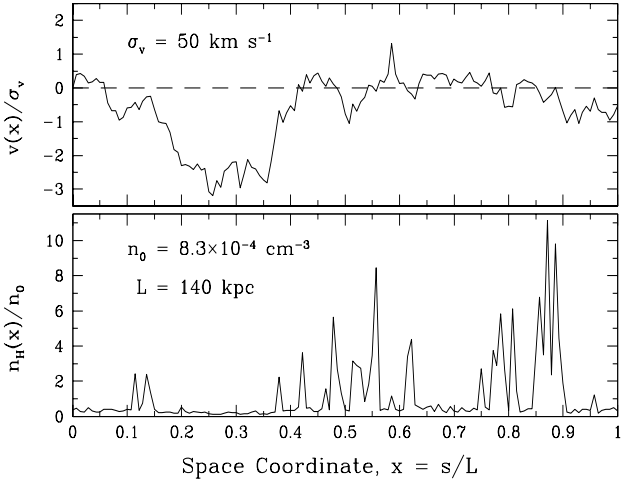,height=8.0cm,width=8.0cm}
\vspace{-1.0cm}
\caption[]{Computed velocity (upper panel) and gas density (lower panel)
distributions along the line of sight within the \zabs = 2.917 absorber
toward HE 0940--1050. Shown are patterns rearranged according to the
principle of minimum entropy production rate (see LACM)
}
\label{fig4}
\end{figure}

Large size, small peculiar velocity dispersion, and
very low metallicity may indicate that the \zabs = 2.917 absorber 
is hosted by either an external galactic halo or by
some large-scale structure object like a filament, 
and may consist of the gas pre-enriched by the first generations of stars.
The sizes of the galactic halos or the filaments have been recently
directly measured at $z < 1$. For instance, observations of \ion{C}{iv}
absorption lines by Chen et al. (2001) revealed that all these lines
can be associated with galaxies seen up to 180 kpc away.   
Penton et al. (2002) reported on large size Ly$\alpha$ absorbers
aligned along a possible filamentary structure of
$\sim 20 h^{-1}_{70}$ Mpc
long and $\sim 1 h^{-1}_{70}$ Mpc wide.

As already mentioned above, the metal content measured in 
this LLS may be considered as primordial, i.e. produced
by the explosions of early generations of stars. Then we may
use the relative metal abundances to restrict masses of these
hypothetical Pop~III objects. Theory predicts
that the abundance ratio [Si/C] rises
to highly supersolar values with increasing masses of the Pop III
stars.
The calculations of the nucleosynthesis patterns of 
metal-free stars with various masses (e.g. Umeda \& Nomoto 2001)  
show that [Si/C] = 0.34, 0.42, 0.70 for, respectively, 
$M/M_\odot = 13, 15, 20$ (Type II SNe); 
[Si/C] = 0.41 -- 0.76 for the Pop III 25$M_\odot$ model with
explosion energies $E_{\rm exp} = 10^{51}-10^{52}$ erg
(Fe Core Collapse Hypernovae); and 
[Si/C] = 1.49 and 1.70 for $M/M_\odot = 170$ and 200, respectively,
in the Pair Instability SNe model\footnote{The [Si/C]
ratios from Umeda \& Nomoto (2001) which 
are normalized to the solar abundance
system given by Anders \& Grevesse (1989) are adjusted to the new solar
photospheric values from Holweger (2001).}.
Similar results are reported in Heger \& Woosley (2002) who also
predicted a pronounced deficit of \ion{Al} compared to \ion{Si}
([Al/Si] $\la -1.5$) in the mass range $140-260 M_\odot$.

In our system we measured [Si/C] = $0.35\pm0.15$ and
[Al/Si] = $-0.18\pm0.13$. These values certainly rule out the
enrichment by very massive stars explosions and restrict the
stellar masses by $\sim 25 M_\odot$.
It should be noted that the abundance ratio [Si/C] $\simeq 0.3-0.4$
is in line with other measurements in high redshift absorbers
(e.g., Songaila \& Cowie 1996; Levshakov et al. 2002b, 2003).
If, nevertheless, the Pop III stars were massive ($M > 140M_\odot$),
then the metallicity produced by their explosions should be much
lower than $10^{-3}Z_\odot$.

We also estimated an upper limit on [N/C] $< -0.4$.     
Nitrogen is usually assumed to be mostly produced by the intermediate
mass stars ($4 \la M/M_\odot \la 8$) whereas carbon production is
dominated by the massive ($M/M_\odot > 8$) stars (Henry et al. 2000).
Unfortunately, the obtained upper limit is too high to allow us to
conclude about the possible nature of nitrogen in the \zabs = 2.917 LLS:
the \ion{N}{iii} $\lambda989$ line lies in the Ly$\alpha$ forest where
the signal-to-noise ratio is only about 25. 

\section{Summary}

We have analyzed the VLT/UVES spectrum of the quasar
HE~0940--1050 and deduced the physical properties of 
the Lyman limit system at \zabs = 2.917.
The main conclusions are as follows~:
\begin{enumerate}
\item  For the first time, a very low carbon abundance of
[C/H] = $-2.93\pm0.13$ is directly measured in the LLS providing
an example of a gas cloud probably enriched by the early generations
($z > 3$) of stars.
\item The analyzed Lyman limit system 
has the line-of-sight size $L$ of about 140 kpc. Its 
velocity dispersion
is only about 50 \kms.
This system may be hosted by
an intergalactic filament structure or an external galactic halo.
\item The measured abundance ratios [Si/C] = $0.35\pm0.15$
and [Al/Si] = $-0.18\pm0.13$ are 
in line with the assumption that
the initial mass functions for early generations of stars
are restricted to intermediate stellar masses, probably
$M_{\rm up} < 25M_\odot$. 
The relative abundances of [Si/C] and [Al/Si] do not reveal any
signature of very massive stars from the early generation.
\item The estimated shape of the photoionizing background radiation
does not confirm that the 
Lyman-continuum radiation
escaping from the young galaxies  
produces 5 times more H-ionizing photons per unit comoving
volume than QSOs at $z \sim 3$ as suggested by Steidel et al. (2001).
None of the far UV galactic spectra calculated by 
Leitherer et al. (1999)
can produce profiles of carbon and silicon ions observed 
in the LLS at \zabs = 2.917. 
\end{enumerate}

\begin{acknowledgements}
We thank Prof. H. Habing for valuable comments
and suggestions.
S.A.L. gratefully acknowledges the hospitality of 
the National Astronomical Observatory
of Japan (Mitaka) where this work was performed. 
The work of S.A.L. and I.I.A. is supported by the 
RFBR grant No.~00-02-16007.
\end{acknowledgements}

\end{document}